\documentclass[floatfix, aps,amsmath,amssymb,superscriptaddress,prb,reprint,twocolumn]{revtex4-2}
\usepackage{graphicx}% Include figure files
\usepackage{float}
\usepackage{dcolumn}% Align table columns on decimal point
\usepackage{multirow}
\usepackage{braket, bm}% bold math
\usepackage[utf8]{inputenc}
\usepackage[T1]{fontenc}
\usepackage{mathptmx}
\usepackage{tabularx}% table size
\usepackage[hidelinks]{hyperref}
\hypersetup{colorlinks, citecolor=blue, urlcolor=blue, linkcolor=blue}
\usepackage{}	% required for `\align' (yatex added)

\begin{document}
\title{Non-Hermitian corner skin effect in a two-dimensional photonic crystal}
\author{Huyen Thanh Phan} 
\email{phanthanhhuyenpth@gmail.com}
\affiliation{Department of Nanotechnology for Sustainable Energy, School
of Science and Technology, Kwansei Gakuin University, Gakuen Uegahara 1, Sanda 669-1330, Japan}
\affiliation{Center for Materials Innovation and Technology, VinUniversity, Hanoi 100000, Vietnam}

\author{Katsunori Wakabayashi}
\email{waka@kwansei.ac.jp}
\affiliation{Department of Nanotechnology for Sustainable Energy, School
of Science and Technology, Kwansei Gakuin University, Gakuen Uegahara 1, Sanda 669-1330, Japan}
\affiliation{Center for Spintronics Research Network (CSRN), Osaka University, Toyonaka 560-8531, Japan}
\affiliation{National Institute for Materials Science (NIMS), Namiki 1-1, Tsukuba 305-0044, Japan}

\date{\today}

\begin{abstract}
We numerically study topological effects of electromagnetic (EM) waves
 in a two-dimensional (2D) non-Hermitian photonic crystal (PhC) composed
 of lossy magneto-optical materials. In this system, not only the EM
 wavefunctions but also the complex eigenfrequencies exhibit nontrivial
 topological properties. We demonstrate that the non-Hermitian skin
 effect, protected by point gaps in the complex eigenfrequency spectrum,
 emerges at both the edges and corners of truncated structures. This
 phenomenon has no counterpart in Hermitian systems. 
In addition, we identify non-Hermitian topological edge states
 originating from the nontrivial topology of the bulk bands. While most
 previous studies of non-Hermitian topology have focused on
 tight-binding models, our work addresses a continuous photonic system,
 providing a more realistic platform and offering a concrete route
 toward experimental realization of non-Hermitian effects. 
\end{abstract} 

\maketitle

\section{Introduction} \label{sec1}
Topological material is a class of quantum materials whose physical
properties are characterized by topological invariants. 
Prominent examples include 
topological insulators~\cite{Fu2007,
Chen2009,Kong2010,Hasan2010,Qi2011,Ando2013}, topological photonic
crystals (PhCs)~\cite{Liu2018,Ozawa2019, Phan2021,Tang2022,Lan2022,
Phan2024} and topological acoustic materials~\cite{Xue2022,Yang2015,
Chen2023,Gao2024}. 
Topology provides a mathematical framework for classifying lattice
systems according to whether their wavefunctions can be continuously
deformed into one another. 
This framework is formalized by topological band
theory~\cite{Bansil2016}, 
which has played a central role in the study of topological materials over the past decades.
According to topological band theory, topological materials are
insulating in the bulk while supporting conducting states at their
boundaries. These boundary states are protected by the nontrivial
topology of the bulk wavefunctions~\cite{Delplace2011,Obana2019,Koizumi2024,Yoshida2019,Kameda2019}. 
This theory was originally developed for Hermitian systems. Recently, it
has been extended to non-Hermitian
systems~\cite{Gong2018,Shen2018,Kawabata2019,Bergholtz2021}, 
where gain and loss naturally arise due to coupling to the
environment. Photonic systems are particularly attractive platforms for
realizing non-Hermitian physics, as gain and loss can be readily
engineered in PhCs~\cite{Teo2022,Wang2023b,Nasari2023}. 

There are several differences between topological band theory for Hermitian and non-Hermitian photonic systems. In Hermitian systems, the emergence of band gap in eigenfrequency spectrum is very important to classify topological properties~\cite{Bansil2016}. On the other hand, the definition of band gap is generalized to point gap, line gap, etc., in non-Hermitian systems~\cite{Gong2018,Kawabata2019,Bergholtz2021,Wang2023b}.
Studies of non-Hermitian topological band theory 
have revealed the non-Hermitian skin effect, in which a macroscopic
number of eigenstates in the finite system become localized at its
boundaries~\cite{Zhong2021,Zhang2020,Okuma2020,Yao2018,Borgnia2020,Zhang2020b}. 
These states are associated with point gaps in the complex eigenfrequency spectrum and are protected by the nontrivial winding of complex eigenfrequencies in the complex plane. This phenomenon is unique to non-Hermitian systems and has no Hermitian counterpart.
On the other hand, 
similar to Hermitian systems, topological properties of bulk wavefunctions in non-Hermitian systems are characterized by complex Berry phase~\cite{Liang2013,Zhao2015,Leykam2017,Pan2018,Dangel2018,Longhi2023}, which can be quantized or non-quantized depending on the symmetries of systems~\cite{Wagner2017,Ding2015,Zhao2015}.

To date, the non-Hermitian skin effect is widely studied in both
tight-binding models and PhCs. However, most of them focus on the
first-order skin effect, where the localization is guaranteed by point
gap of complex bulk eigenfrequencies. Studies on the second-order skin
effect, where the localization emerges at corner of finite structures,
are still limited. Some studies proposed the second-order skin effect
based on Benalcazar-Bernevig-Hughes model
\cite{Okugawa2020,Kawabata2020}, where positive/negative hoppings
parameters are required for forward/backward directions. 
This model has not yet been realized experimentally.
Besides, corner skin modes are
numerically observed in a two-dimensional (2D) PhC with gain and loss
under magnetic field~\cite{Zhu2023,Liu2024}. 
To experimentally realize non-Hermitian effects, further materials design and theoretical
calculations are required. 

In this paper, we theoretically design a 2D non-Hermitian photonic
crystal made of magneto-optical materials
that break inversion, mirror, and time-reversal symmetries.
The first-order
non-Hermitian skin effect is numerically observed, which is protected by
non-trivial winding of the complex bulk eigenfrequencies. Furthermore,
non-Hermitian topological edge states are also obtained in the line gap
between the first and the second bands. We numerically calculate the
complex topological phases of the bulk wavefunctions, which is
consistent with the emergence of topological edge states. Complex
eigenfrequencies of these topological edge states also form point gap in
the complex plane. This point gap is responsible for the emergence of
the second-order non-Hermitian skin effect at the corner of finite
structure. 
Our results provide essential insights into non-Hermitian topological
states and represent a step toward the experimental realization of
non-Hermitian effects. 

\section{Theoretical background} \label{section2}
Before studying non-Hermitian properties of a 2D PhC system, in this section, we give a brief review of theoretical background. In general, eigenfrequencies of a non-Hermitian system are complex numbers. Assuming the lattice constant is $a=1$, there are two possible cases of one-dimensional (1D) non-Hermitian band structure~\cite{Zhong2021}. The first case is called reciprocal band structure, where $\omega(k) = \omega(-k)$ as shown in Fig. 1(a). This band structure becomes a curved line, which is often called trivial loop in the complex plane as denoted in Fig. 1(b). On the other hand, when $\omega(k) \neq \omega(-k)$, band structure will form a closed loop (nontrivial loop) in the complex plane as shown in Figs. 1(c) and (d). The area inside the closed loop is called point gap.
To analyze properties of the point gap, the winding number is involved and defined as~\cite{Zhong2021}
\begin{equation} \label{winding}
   w = \frac{1}{2\pi}\int_{-\pi}^{\pi} \partial_k \text{arg}(\omega(k)-\omega_0) dk,
\end{equation}
where $\omega$ denotes the complex eigenfrequency of a given band, $k$
is wave vector, $\omega_0$ is any frequency in the complex plane. 

\begin{figure}[ht!]
\centering\includegraphics[width=0.4\textwidth]{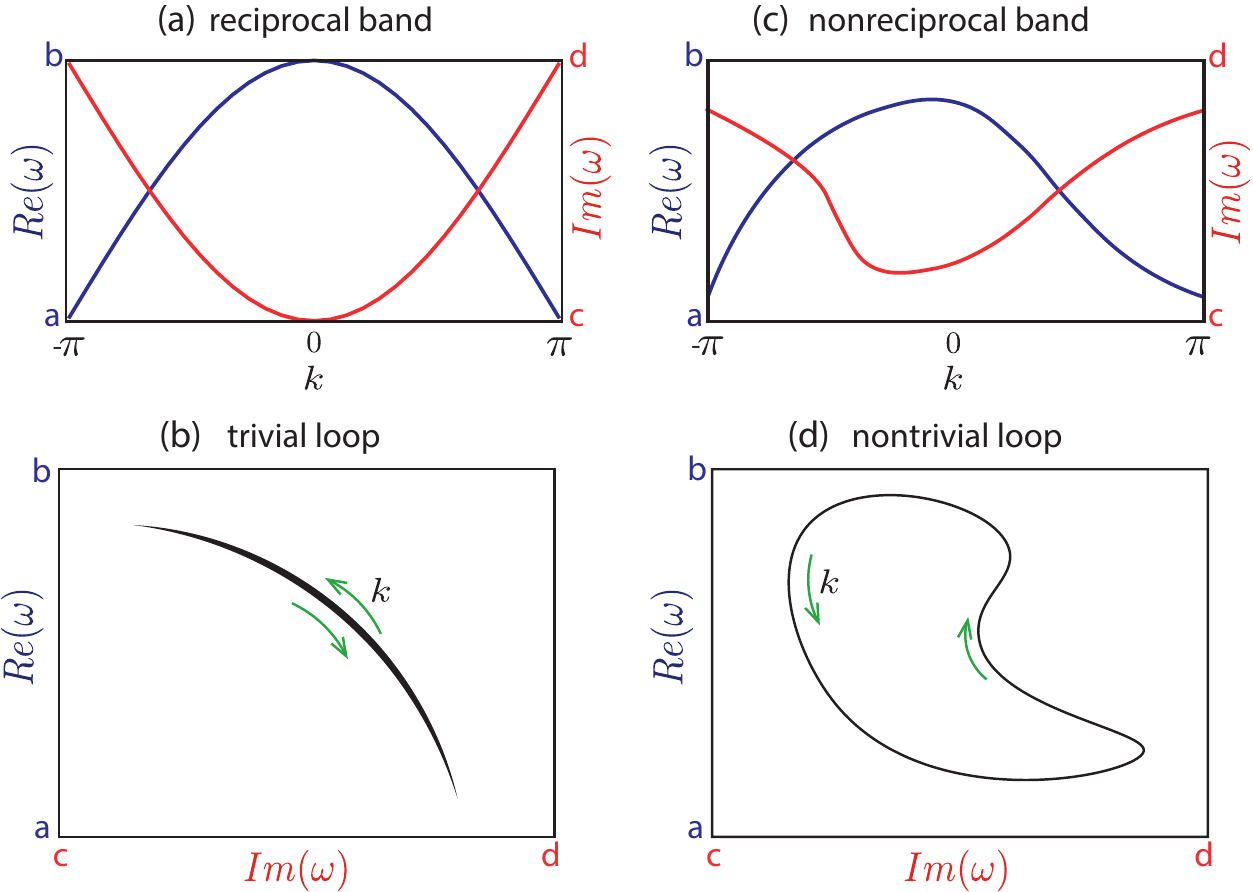}
\caption{(a) Schematic of a 1D reciprocal complex band structure, where $\omega(k) = \omega(-k)$. (b) The complex eigenfrequencies in (a) form a curved line (trivial loop) in complex plane. (c) Schematic of a 1D nonreciprocal complex band structure, where $\omega(k) \neq \omega(-k)$. (d) The complex eigenfrequencies in (c) show a closed loop (nontrivial loop) in complex plane.} 
\label{figure1}
\end{figure}

In Eq.~(\ref{winding}), the winding number $w$ is the number of times that complex eigenfrequencies wind around point $\omega_0$ when $k$ is varied in the first Brillouin zone (BZ).
For any point inside the point gap, the winding number is nonzero. 
According to the above analysis, the necessary condition for obtaining a nontrivial point gap in 1D system is $\omega(k) \neq \omega(-k)$. This corresponds to the broken inversion symmetry in real space and broken reciprocity~\cite{Zhong2021}.
For 2D systems, where the angular frequencies are defined by $\omega(k_x,k_y)$, the integration in Eq.~(\ref{winding}) is taken along one direction in k-space, the winding number will depend on the k-points in the other direction. For example, the $k_y$-dependent winding number is defined as follow
\begin{equation}\label{winding2D}
    w(k_y) = \frac{1}{2\pi}\int_{-\pi}^{\pi} \partial_{k_x} \text{arg}(\omega(k_x, k_y)-\omega_0) dk_x.
\end{equation}
In this case, to achieve non-trivial loop, the necessary condition is $\omega(k_x,k_y) \neq \omega(-k_x,k_y)$. This means that the mirror symmetry with respect to $k_x=0$ axis and/or the reciprocity of system are broken~\cite{Zhong2021}.

It is mentioned in several studies that the non-zero winding number of
the bulk band structures are responsible for high concentration of EM
waves at the edges of truncated
structures~\cite{Gong2018,Yao2018,Zhang2020b,Okuma2020}. In particular,
if complex eigenfrequencies of infinite periodic systems form point gap
in the complex plane, those of their truncated structures, which have at
least one open boundary, will be within the point gap. 
Their field distribution are concentrated at the open boundaries and
decay to the bulk region. These phenomena are called non-Hermitian skin
effect, which is unique in non-Hermitian systems without Hermitian
counterpart. 
To elucidate the origin of this effect, note that when the complex eigenfrequencies exhibit point gaps (Fig.~\ref{figure1}(d)), a single real eigenfrequency corresponds to two distinct imaginary eigenfrequencies, which in turn correspond to two $k$ values with opposite signs. This means that, at the open boundaries, the incoming waves and reflected waves have different decay rates. When the reflected waves decay more slowly than the incoming waves, the skin effect emerges because residual energy accumulates at the boundary. On the other hand, when no point gap is present (Fig.~\ref{figure1}(c)), each real eigenfrequency is associated with a single imaginary eigenfrequency. This complex eigenfrequency then maps to two opposite $k$ values, resulting in identical decay rates for the incoming and reflected waves and thus no skin effect is observed.

\section{The first-order non-Hermitian Skin Effect and topological edge states}\label{section3}

\begin{figure}[ht!]
    \centering
    \includegraphics[width=0.48\textwidth]{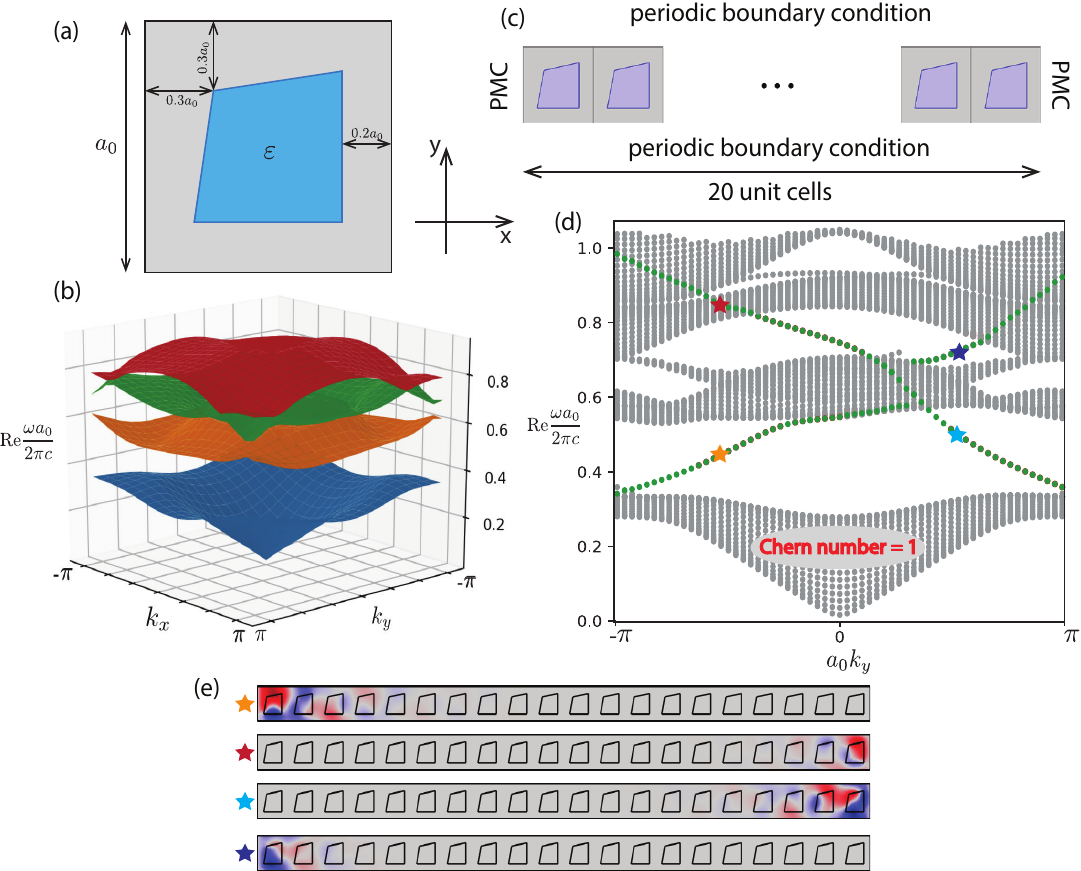}
    \caption{(a)Schematic of one unit cell of investigated 2D non-Hermitian PhC. (b) $k$-dependent dispersion of real part of eigenfrequencies, showing a complete band gap between the first and the second bands. (c) Schematic of a super cell for ribbon structure, which is periodic along $y$ direction and finite in $x$ direction. (d) Photonic band structure for the structure in (c). Topological edge states (green color) are numerically observed above the first bulk band. (e) Field profile of the topological edge states marked by colored stars in (d).} 
    \label{figure2}
\end{figure}

In this section, based on the theoretical background in
section~\ref{section2}, we will design a 2D non-Hermitian PhC, 
which exhibits the non-Hermitian skin effect at both 1D edges and 0D corners.
Our PhC structure is made of magneto-optical materials, where non-symmetric blocks are arranged periodically in air medium.
Figure~\ref{figure2}(a) is the schematic of a square unit cell with
lattice constant is $a_0$. Inversion symmetry and mirror symmetry with
respect to $x$ and $y$ direction are broken in this PhC. We assume that
the magnetization of materials is parallel to the $z$ direction. Therefore, its relative permittivity is~\cite{Zvezdin1997, Kotov2018}
\begin{equation}\label{relaper}
\varepsilon = 
\begin{bmatrix}
  4.2-0.6i & -5i & 0 \\
  5i & 4.2-0.6i & 0 \\
  0 & 0 & 1
\end{bmatrix},
\end{equation}

In Eq.~(\ref{relaper}), the diagonal terms represent the actual complex relative permittivity of the materials, while the off-diagonal terms denote the strength of the nonreciprocity. We note here that, the main results are robust over the broad parameter range, as long as the point gap is preserved. 

\begin{figure*}[ht!]
    \centering
    \includegraphics[width=0.95\textwidth]{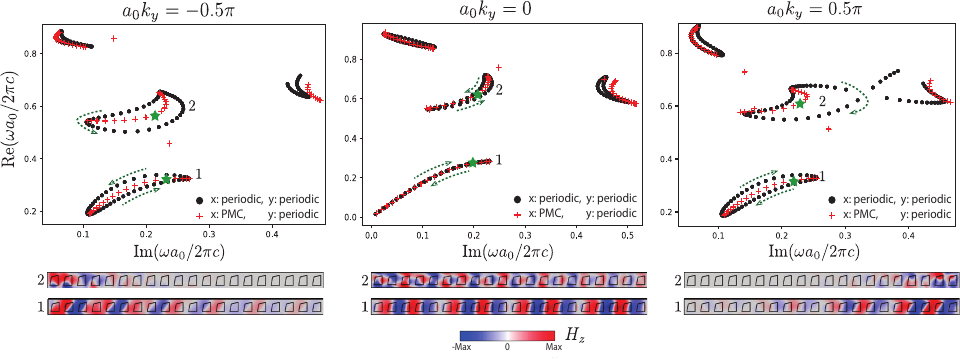}
    \caption{Photonic band structure for the infinite structure (black) and ribbon structure (red) at $a_0k_y = -0.5\pi$ (left panel), $a_0k_y = 0$ (middle panel) and $a_0k_y = 0.5\pi$ (right panel). The green arrows denote the winding direction (clockwise or anti-clockwise) when $k_x$ is varied from $-\pi/a_0$ to $\pi/a_0$. Number 1 and 2 indicate band indices.The lower panels are field profile of the states labeled by green stars in the corresponding band structures. The states enclosed by the point gap exhibit skin effect, where EM waves are highly localized at the boundaries of structures.}
    \label{figure3}
\end{figure*}

To numerically determine photonic band structure and topological properties of this non-Hermitian PhC, we use the finite-element method from COMSOL Multiphysics and the extended plane-wave expansion method~\cite{Hsue2005}.
Throughout this paper, the transverse electric (TE) waves, where
electric field is parallel to the $xy$ plane and the magnetic field is perpendicular to $xy$ plane, will be considered. The dependence of the real part of eigenfrequencies on wave vectors ${\bf k}$ is shown in Fig.~\ref{figure2}(b). There is a band gap between the first and the second lowest bands. 
In Fig.~\ref{figure2}(c), we show a super cell consisting of 20 and 1 unit cell in $x$ and $y$ directions, respectively. In order to examine ribbon structure, the periodic boundary condition is applied to the upper and lower boundaries, the perfect magnetic conductor (PMC) boundary condition is imposed on the left and right boundaries. Photonic band structure for the ribbon is shown in Fig.~\ref{figure2}(d) as the $k_y$-dependence of the real part of eigenfrequencies. Since the permittivity tensor is asymmetric, system becomes nonreciprocal, leading to the non-equivalence of two boundaries on the left and right sides of the ribbon structure. Therefore, two separated edge states colored in green are numerically observed above the first bulk band, where one state is the edge states on the left side and the other one is the edge state on the right side as shown in Fig.~\ref{figure2}(e).
Origin of these two edge states will be studied in the next section.

In Fig.~\ref{figure3}, we extract the complex eigenfrequencies of both infinite and ribbon structures at several fixed values of $k_y$ and plot them in the same complex plane. For the infinite structure, which is periodic in both the $x$ and $y$ directions, the eigenfrequencies (black dots) at one fixed $k_y$ and $k_x \in [-\pi/a_0, \pi/a_0]$ form closed loops. On the other hand, the eigenfrequencies of the ribbon structure, subject to a PMC boundary condition on the left and right boundaries and periodic boundary conditions along the $y$ direction, are indicated by red plus signs, which form arcs. It is evident that the eigenfrequency spectrum of the ribbon structure lies predominantly within the region enclosed by that of the infinite structure. As discussed in section~\ref{section2}, any point located inside the point gap is associated with a nonzero winding number. Therefore, any eigenfrequencies of semi-infinite structure that lie within the loops will possess a nonzero winding number. These states exhibit skin effect, where EM waves are highly localized at the boundaries of the ribbon structures as shown in the bottom panels of Fig.~\ref{figure3}. 
The skin effect can be described within non-Bloch framework by complexifying the wave vectors; see Appendix A for details.
From the topological point of view, the non-Hermitian skin effect is
topologically protected by nonzero winding number. 
This skin effect may emerge at the left or right boundaries as shown in
the field profile in Fig.~\ref{figure3}, which depends on the sign of
winding numbers. For example, at $a_0k_y = -0.5\pi$, the winding number
for both the first and second bands is $1$, skin effect emerges at the
left boundary. Meanwhile, the winding number for both the first and 
second bands at $a_0k_y = 0.5\pi$ is $-1$, skin effect occurs at the
right boundary. The winding number of point-gap is generally
$k_y$-dependent except only some certain intervals such as $k_y = 0, \pm
\pi/a_0$. Their sign flip is reflected by the sign change of the complex
geometrical phase as shown in the next section and in
Fig.~\ref{figure4}(a).  

In Fig.~\ref{figure3}, we present the complex eigenfrequency spectra
under periodic boundary conditions along the y-direction. Identical
spectra are obtained when periodic boundary conditions are applied along
the x-direction. There are, however, differences in the spatial
localization of the skin effect. Specifically, if the skin effect
appears at the left (right) boundaries when the system is truncated
along the y-direction, the corresponding localization shifts to the
upper (lower) boundaries when the system is truncated along the
x-direction. 

We note here that the skin effect in non-Hermitian PhCs is the effect
where a macroscopic number of bulk eigenstates become localized and decay
 exponentially at the boundaries under open boundary conditions. The
 eigenfrequencies of the skin modes are within the bulk regions. In
 addition with the skin modes, there is also other edge states with the
 eigenfrequencies in the gap regions as shown in
 Fig.~\ref{figure2}(d). These edge states may originate from nontrivial
 topology of the bulk as the non-Hermitian counterpart of the Hermitian
 topological edge states. To confirm this statement, we will numerically
 calculate topological phase of the first bulk band.

\section{Non-Hermitian geometrical phase} \label{section4}

Topological phase of 2D systems is characterized by the Zak
phase~\cite{Delplace2011} or the Wilson loop~\cite{Wang_2019}, which is
the geometrical phase acquired by eigenstate when it is transported
along a path of BZ. In the Hermitian band theory, the Berry connection
is defined as 
\begin{equation} \label{berryconnection}
    {\bf A}_{n{\bf k}} = i\langle {\bf u}_{n{\bf k}}| \partial_{\bf k}|{\bf u}_{n{\bf k}}\rangle,
\end{equation}
where ${\bf u}_{n\bf k}$ is the periodic part of the Bloch wave functions ${\bf H}_n({\bf r}) = {\bf u}_{n\bf k}({\bf r}) e^{i{\bf k} \cdot {\bf r}}$, $n$ is band index, ${\bf k} = (k_x, k_y)$. In Eq.~(\ref{berryconnection}), the bra $\langle {\bf u}_{n\bf k}|$ and the ket $|{\bf u}_{n\bf k}\rangle$ are the complex conjugate of each other. Also, they are the left and right eigenvectors of the eigenvalue equation, respectively. If systems are non-Hermitian, the Eq.~(\ref{berryconnection}) becomes ambiguous since the left and right eigenvectors of the eigenvalue equation are distinct~\cite{Wang2023b}. In non-Hermitian systems, the Berry connection is defined as~\cite{Fan2020, Wang2023b}
\begin{equation} \label{berryconnection2}
    {\bf A}^{\alpha \beta}_{n{\bf k}} = i\langle {\bf u}^{\alpha}_{n{\bf k}}| \partial_{\bf k}|{\bf u}^{\beta}_{n{\bf k}}\rangle,
\end{equation}
where $\alpha, \beta \in {L,R}$ denote left ($L$) or right ($R$) eigenvectors. Since $\langle {\bf u}^{\alpha}_{n\bf k}|$ and $|{\bf u}^{\alpha}_{n\bf k}\rangle$ are not the complex conjugate of each other in general, the Berry connection becomes complex. Therefore, the total Berry phase (geometrical phase) is also complex, which is calculated by integrating the Berry connection over a closed path (L) of the first BZ.
\begin{equation} \label{berryphase}
    \gamma^{\alpha \beta} = \oint_{\text{L}} {\bf A}^{\alpha \beta}_{n{\bf k}} d{\text{L}},
\end{equation}

In our study, we create a ribbon structure, which is periodic in $y$ direction and finite in $x$ direction. The isolated edge states in the band gap are paralleled to the $y$ direction. Therefore, we will calculate the Berry phase in $y$ direction (often-called the Zak phase) as follow
\begin{equation} \label{berryphase2}
    \gamma_n^{LR}(k_y) = \int_{-\pi/a_0}^{\pi/a_0} i\langle {\bf u}^{L}_{n{\bf k}}| \partial_{k_x}|{\bf u}^{R}_{n{\bf k}}\rangle dk_x,
\end{equation}

\begin{figure}[ht!]
\centering\includegraphics[width=0.45\textwidth]{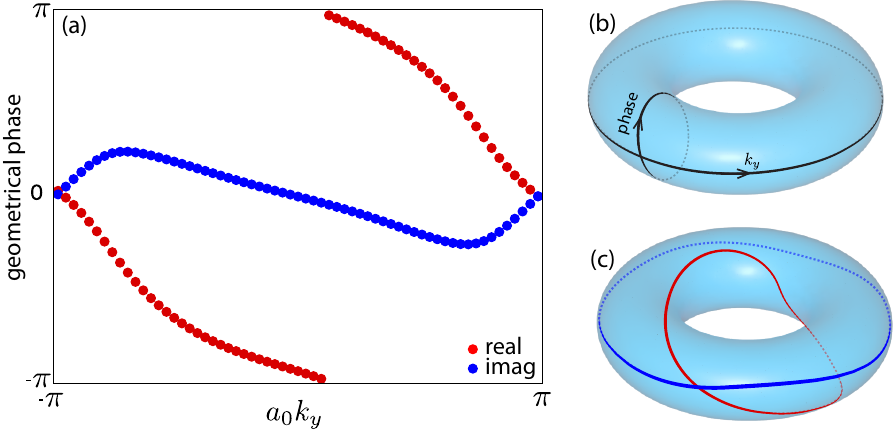}
\caption{(a) Complex geometrical phase in $y$ direction of the first band of investigated 2D non-Hermitian PhC. Red dots indicate real part of geometrical phase, blue dots denote imaginary part of geometrical phase. (b) The coordinate system in (a) is equivalent to a torus under periodic boundary condition. (c) The real and imaginary part of the complex geometrical phase are represented by red and blue lines on the torus.}
\label{figure4}
\end{figure}

Applying the numerical calculation method and the gauge smoothing process described briefly in Appendix B and detailedly in ref.~\cite{Wagner2017}, we obtained the complex geometrical phase in $y$ direction for the first lowest band as shown in Fig.~\ref{figure4}(a). The red and blue dot lines indicate the real and imaginary part of geometrical phase, respectively. It is the fact that the first point ($a_0k_y = -\pi$) and the last point ($a_0k_y = \pi$) of the first BZ are equivalent, the value $\pi$ and $-\pi$ of the geometrical phase are also equivalent. Therefore, the coordinate system in Fig.~\ref{figure4}(a) is equivalent to a torus as shown in Fig.~\ref{figure4}(b) under periodic boundary condition for the Bloch states. We define the winding of geometrical phase as
\begin{equation}
    \label{winding2}
    \text{w}_n = \frac{1}{2\pi}\oint_{k_y} \partial_{k_y} \gamma_n^{LR}(k_y) dk_y, 
\end{equation}

By using Eq.~(\ref{winding2}), we obtained the winding number for the real and imaginary part of complex geometrical phase is $1$ and $0$, respectively. We describe the winding properties of the complex geometrical phase on the torus as shown in Fig.~~\ref{figure4}(c). While the real part (red line) winds once along the torus, the imaginary part (blue line) changes smoothly without any winding. Since the Chern number is consistent with this winding number~\cite{Weng2015,WindingChern}, the Chern number for the first lowest band of our investigated PhC is 1. This is responsible for the emergence of edge states above the first bulk band shown in Fig.~\ref{figure2}(d). 

\section{Corner Skin Effect} \label{section5}

\begin{figure*}[ht!]
\centering\includegraphics[width=0.8\textwidth]{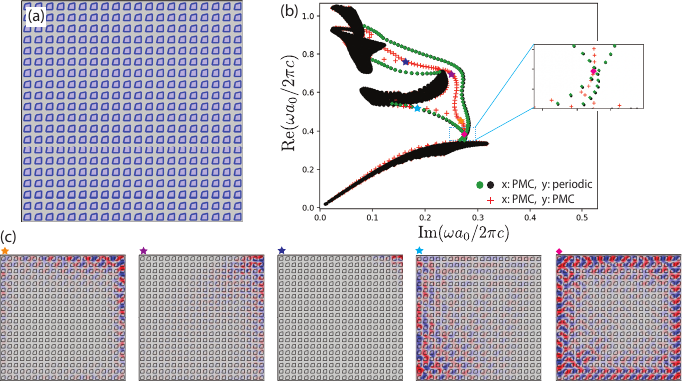}
\caption{(a) Schematic of investigated finite structure, system size is $20\times 20$ unit cells. PMC boundary condition is applied to all four boundaries. (b) Photonic dispersion in the complex plane of the finite structure (red) in (a) and semi-infinite (black) in Fig.~\ref{figure2}(c). The inset shows enlarged figure of the region enclosed by the blue dot line. The topological edge states form loop in complex plane and encloses the complex eigenfrequencies of the corner modes. (c) Field profile of the corner skin modes labeled by the stars in (b) and the extended topological edge states labeled by the magenta circle in (b).}
\label{figure5}
\end{figure*}

In this section, we numerically study the second-order skin effect. A finite structure which consist of $20 \times 20$ unit cells with PMC boundary condition in both $x$ and $y$ directions are investigated as shown in Fig.~\ref{figure5}(a). For the ribbon structure in Fig.~\ref{figure2}(c), its photonic dispersion are indicated by the black dots in Fig.~\ref{figure5}(b). In the inset, we enlarge the region that is closed to the first band. There is a crossing point of the topological edge states in this region.
It is easily seen that the topological edge states form point gap in the complex plane. 

The photonic dispersion of finite structure in Fig.~\ref{figure5}(a) is
plotted as red plus sign in Fig.~\ref{figure5}(b). The states within the
point gap exhibit corner modes, where EM waves are highly localized at
the corners and decay exponentially. It can also be understood that
all of the topological edge states of the semi-infinite structure are
collapsed to corner skin modes when the structure is truncated in both
directions.  
It is evident that the top (bottom) and left (right) boundaries share the same geometrical structure.
Although PMC boundary conditions are imposed on all boundaries of the finite structure, the point gap arises from topological edge states associated with two distinct boundary geometries. 
Therefore, corner modes are only observed at the top right and bottom left corners as shown in the first four panels of Fig.~\ref{figure5}(c). At the frequency of the crossing point, the extended topological edge states happen as shown in the last panel of Fig.~\ref{figure5}(c). This exception arises because at the crossing point, one state corresponds to the positive-$k$ mode localized at the right (bottom) boundary, while the other corresponds to the negative-$k$ mode localized at the top (left) boundary. Consequently, at the corner, the incoming and reflected waves have identical decay rates, thereby canceling the skin effect.

To distinguish from the topological corner states in Hermitian systems,
these corner states are called corner skin states, which are protected
by the point gap of the investigated photonic system. This means that
whenever the point gap emerges among topological edge states, the corner
skin effect happens. While the number of topological corner states is
limited and depends on the number of corners in each specific structure, the
number of corner skin modes is much larger than that of topological corner state and depends on the system size. 
We conclude that the emergence of point gap in the edge-state spectrum is the condition for the corner skin effect, as double truncation causes these edge states to collapse and accumulate at the corner.

\section{Summary and conclusions}
In summary, we have theoretically designed a 2D
non-Hermitian photonic crystal and numerically investigated the skin
effect and topological states of electromagnetic waves. Owing to the
absence of inversion and mirror symmetries as well as broken
reciprocity, complex eigenfrequencies exhibit point gaps in the complex
plane, which protect the emergence of the non-Hermitian skin effect at
the edges of truncated structures. 

Furthermore, we demonstrated that topological edge states arising from a
nonzero Chern number also form point gaps in their complex
eigenfrequency spectra. These point gaps give rise to the second-order
non-Hermitian skin effect, where electromagnetic waves become localized
at the corners of finite structures. Unlike conventional topological
edge or corner states, the skin modes appear over broad frequency
range and scale with system size. Our results provide a realistic
photonic platform for exploring higher-order non-Hermitian topology and
may stimulate future experimental studies.

%We have theoretically designed a 2D non-Hermitian PhC and numerically examined skin effect and topological states of EM waves propagating in it. Based on the theoretical background, our PhC structure is not only in the absence of mirror and inversion symmetries but the reciprocity is also broken. The point gaps of the complex eigenfrequencies in the complex plane are numerically observed, which protect the emergence of skin effect at the edge of truncated structure. Moreover, the topological edge states, which originates from Chern number, are also obtained in the real band gap. The complex eigenfrequencies of these topological edge states also form point gap in the complex plane. This point gap protects the emergence of corner skin effect, where EM waves are localized at the corner of a finite structure. The extended topological edge states are also observed in this finite structure. 
%
%Our finding of EM waves localization at the edges and corners due to non-Hermitian skin effect could motivate the future experimental interest. The reason for this statement is that edge and corner skin modes are observed over broad range of frequency, rather than being restricted to the narrow range as topological edge states or just single frequency as topological corner states. This property are applicable to the enhancement of the light-matter interaction. 
%

\section*{Acknowledgements}

This work was supported by JSPS KAKENHI (Grants No. JP25K01609, No. JP22H05473, and No. JP21H01019), JST CREST (Grant No. JPMJCR19T1). K.W. acknowledges the financial support for Basic Science Research Projects (Grant No. 2401203) from the Sumitomo Foundation.

\section*{DATA AVAILABILITY}
The data used and analyzed during the current study available from corresponding authors on reasonable request.

\section*{Appendix A: Using complex wave vectors to describe skin modes}
\label{appendixA}

In this Appendix, we present a practical method for obtaining complex wave vectors (a non-Bloch implementation) in 2D continuous photonic systems.
EM waves propagating in PhC systems are mathematically described by the
Bloch band theory, which includes a periodic term and a plane wave term
with real wave vectors ${\bf H}_n({\bf r})  = {\bf u}_{n\bf k}({\bf r})
e^{i{\bf k \cdot r}}$. 
Following Bloch band theory, the band structure
of a finite large system can be reproduced by using one unit cell with
periodic boundary condition. However, in a truncated non-Hermitian
system, 
a macroscopic number of bulk states are not extended throughout the entire structure.
They become localized at the boundaries and exponentially
decay.  
Moreover, it is easily seen in Fig.~\ref{figure3} that the eigenfrequencies of one unit cell with periodic boundary condition (black) and the eigenfrequencies of the structure with two PMC boundaries (red) are distinct. Therefore, the conventional Bloch band theory with real wave vectors ${\bf k}$ is not suitable to describe the skin modes. To ensure the mathematical formulation accurately captures the physics, the localization and decay properties of skin effect will be expressed using complex wave vectors ${\bf\tilde{k}}$ instead of real ones. This framework corresponds to the non-Bloch band theory~\cite{Yokomizo2019, Yokomizo2023}.

\begin{figure}[ht!]
\centering\includegraphics[width=0.4\textwidth]{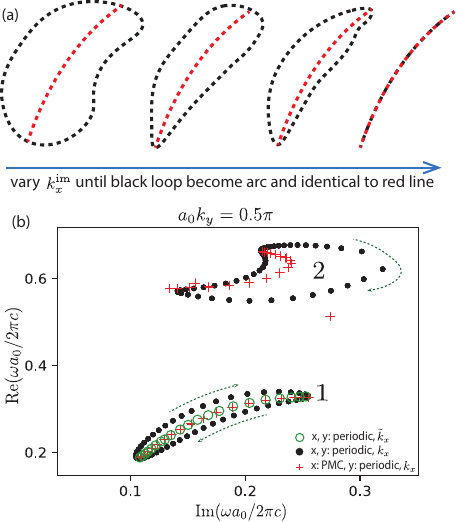}
\caption{(a) Schematic of eigenfrequencies of the infinite structure (black) and eigenfrequencies of semi-infinite structure (red). From left to right panels, $k_x^{\text{im}}$ is varied from 0 to the critical point, where eigenfrequencies of two structures become identical. (b) Band structure of infinite structure with real $k_x$ (black), semi-infinite structure (red) and infinite structure with complex $\tilde{k}_x$.}
\label{figure6}
\end{figure}

In our investigated ribbon structure, the skin effect emerges at the two
boundaries paralleled to $y$ direction. EM waves decay in $x$
direction and extend in $y$ direction. Therefore, we assume that $k_x$
is complex $\tilde{k}_x = k_x^{\text{re}} + ik_x^{\text{im}}$ and $k_y$
is real as normal. We note here that both $k_x^{\text{re}}$ and $k_y$
take values from $-\pi/a_0$ to $\pi/a_0$ in the first BZ. To find
$k_x^{\text{im}}$, the transfer matrix method or the tight binding
approximation method can be used. These two methods work well for 1D
systems~\cite{Yokomizo2022} and $k_x^{\text{im}}$ can be solved
analytically. However, for 2D systems, these two methods become
complicated since there are several dependent parameters. As mentioned
previously, we expect to reproduce the photonic band structure of
truncated systems using one unit cell with periodic boundary
condition. Therefore, here, we introduce a simulation method, which can
be used to obtain $k_x^{\text{im}}$ more easily than tight binding
approximation and transfer matrix method. 

In Fig.~\ref{figure6}(a), the black and red dashed lines denote the eigenfrequencies of the infinite and ribbon structures, respectively. The red lines remain fixed throughout the procedure. For the infinite structure, the eigenfrequencies are recalculated by employing complex wave vectors $\tilde{k}_x$. The imaginary component $k_x^{\text{im}}$ is systematically varied until the eigenfrequencies of the infinite (black) and ribbon (red) structures become identical. The corresponding value of $k_x^{\text{im}}$ is thereby identified as the desired one.

Focusing on the lowest photonic band at $a_0k_y = 0.5\pi$ as shown in Fig.~\ref{figure6}(b), the point gap form by eigenfrequencies of infinite structure (black) is collapsed to an arc (green) when the imaginary component $k_x^{\text{im}}$ is $0.048\pi/a_0$. This arc coincides with the eigenfrequencies of the ribbon structure (red). This means that the eigenfrequencies of skin modes are reproduced by using one unit cell with periodic boundary condition and complex wave vectors. The localization length of the wavefunctions are determined by the imaginary part of complex wave vectors, given by $1/|k_x^{\text{im}}|$. In this case, at $a_0k_y = 0.5\pi$, localization length is $1/|k_x^{\text{im}}| = 1/|0.048\pi/a_0| \approx 7a_0$. This value is consistent with the field profile shown in the left panel of Fig.~\ref{figure3}.

\section*{Appendix B: Numerical gauge smoothing for the complex Berry phase} \label{appendixB}

In this appendix, we give a brief review of a numerical method to obtain the left and right eigenvectors $\langle {\bf u}^{L}_{n{\bf k}}|$ and $|{\bf u}^{R}_{n{\bf k}}\rangle$ and in a suitably smooth gauge, enabling the calculation of complex Berry phases along a discretized $k_x$ in parameter space~\cite{Wagner2017}. This procedure is essential for the numerical evaluation of Eq.~(\ref{berryphase2}). The properties of EM waves in photonic crystals are analyzed using the following eigenvalue equation
\begin{equation}
    \nabla \times \frac{1}{\varepsilon(\bf r)} \nabla \times {\bf H(r)}
    = \left(\frac{\omega}{c}\right)^2 {\bf H(r)},
\end{equation}
here, we put $\Theta = \nabla \times \frac{1}{\varepsilon(\bf r)} \nabla \times$ is the operator. The left and right eigenvectors can be obtained by 
\begin{subequations}
\begin{align}
    \langle {\bf u}_{L}^n ({\bf k})| \Theta &= \langle {\bf u}_{L}^n ({\bf k})| \left(\frac{\omega}{c}\right)^2, \\
    \Theta |{\bf u}_{R}^n ({\bf k})\rangle &= \left(\frac{\omega}{c}\right)^2 |{\bf u}_{R}^n ({\bf k})\rangle.
\end{align}
\end{subequations}

In general, the left and right eigenvectors are computed independently for each discrete $k$-point, and each carries an arbitrary global phase. According to ref.~\cite{Brody2014}, the biorthogonal
normalization condition for the left and right eigenvectors at each $k_i$ point in $k_x$ direction  are

\begin{subequations}
\label{normalization}
\begin{align}
    \langle {\bf u}_{L}^n (k_i)| &\rightarrow \frac{\langle {\bf u}_{L}^n (k_i)|}{\sqrt{\langle {\bf u}_{L}^n (k_i)|{\bf u}_{R}^n (k_i)\rangle}}, \\
    |{\bf u}_{R}^n (k_i)\rangle  &\rightarrow \frac{|{\bf u}_{R}^n (k_i)\rangle} {\sqrt{\langle {\bf u}_{L}^n (k_i)|{\bf u}_{R}^n (k_i)\rangle}}.
\end{align}
\end{subequations}

By normalizing left and right eigenvectors, $\langle {\bf u}_{L}^m (k_i)|{\bf u}_{R}^n (k_i)\rangle = \delta_{mn}$ is satisfied. To smooth the gauge in the 1st BZ, we choose a based point such as $k_1$, then normalized the eigenvectors according to Eq.~(\ref{normalization}). The arbitrary global phases are adjusted by the following process
\begin{subequations}
    \begin{align}
        \langle {\bf u}_{L}^n (k_i)| &\rightarrow \langle {\bf u}_{L}^n (k_i)| e^{-i \text{arg} \langle {\bf u}_{L}^n (k_i)|{\bf u}_{R}^n (k_{i-1})\rangle}, \\
        |{\bf u}_{R}^n (k_i)\rangle &\rightarrow |{\bf u}_{R}^n (k_i)\rangle e^{-i \text{arg} \langle {\bf u}_{L}^n (k_{i-1})|{\bf u}_{R}^n (k_i)\rangle}.
    \end{align} 
\end{subequations}

After the above step, the gauge is basically smooth. However, they are not single-valued in the 1st BZ. In particular, at the starting point $k_1$ and ending point $k_N$, eigenvectors are not identical. Next, we adjust the phase difference between the starting and ending point. The argument of the first non-vanishing component $p$ of the left eigenvector $\langle {\bf u}_{L}^n (k_i)|$ is $\phi_i$. We define a continuous function as follow
\begin{equation}
    f(\Delta \phi, k_x) = \Delta \phi \frac{k_x + \pi/a}{2\pi/a},
\end{equation}
where $\Delta \phi = \sum_{i=1}^{N-1} (\phi _{i+1} - \phi_i)$ is the phase difference between $k_1$ and $k_N$. The final phase adjustment step is 
\begin{subequations}
    \begin{align}
        \langle {\bf u}_{L}^n (k_i)| &\rightarrow \langle {\bf u}_{L}^n (k_i)| e^{-i f(\Delta \phi, k_i)}, \\
        |{\bf u}_{R}^n (k_i)\rangle  &\rightarrow |{\bf u}_{R}^n (k_i)\rangle e^{i f(\Delta \phi, k_i)}.
    \end{align}
\end{subequations}

Up to here, gauge is smooth and the biorthogonal normalization $\langle {\bf u}_{L}^m (k_i)|{\bf u}_{R}^n (k_i)\rangle = \delta_{mn}$ is still ensured.  

% \nocite{*}
% \bibliography{apssamp}% Produces the bibliography via BibTeX. 

%\bibliographystyle{apsrev4-2}
\bibliography{references}
\end{document}